# Colinear ternary fission and nucleon phase model

## G. Mouze, S. Hachem and C. Ythier

Faculté des Sciences, Université de Nice, 06108 Nice cedex 2, France

mouze@unice.fr

**Abstract:** The colinear ternary fission mode discovered by Pyatkov et al. can be interpreted in the framework of the nucleon-phase model of fission.

**PACS:**

25.85.-w: Fission reactions;

21.60 Gx :Cluster models

## 1. Introduction

The existence of a new mode of ternary nuclear fission was recently suggested by Yu.V. Pyatkov et al. [1,2]. According to these authors, this mode occurs in the spontaneous fission of $^{252}$Cf and in the neutron-induced fission of $^{235}$U with a relatively high yield of the order of 4-5 $10^{-3}$ per binary fission; the new phenomenon could be understood if one assumes a collective motion through hyper- deformed shapes of the fissioning system; in support of this assumption, they point out that the linear arrangement of the fragments provides the lowest Coulomb potential energy.

The aim of the present paper is to show that a colinear emission of three fission fragments can be predicted by the nucleon-phase model of binary nuclear fission [3-6].

## 2. The nucleon-phase model of nuclear fission

According to this model, nuclear fission occurs in three steps. In its first step, the fissioning systems clusterize into a dinuclear system made of a $^{208}$Pb core and a light cluster, e.g.:

$$^{239}\text{Pu} + \text{n} \rightarrow {}^{208}\text{Pb} + {}^{32}\text{Mg} + 79.36 \text{ MeV } [7] \tag{1}$$

In the second step, the collision of core and cluster leads to the destruction of the $^{208}$Pb core, to the creation of a hard A =126- nucleon core, and to the release of <u>82</u> nucleons. But a number of these nucleons are captured by the cluster in order to form a hard A = 82-nucleon core. The asymmetric mass distribution results from the sharing- out of the remaining <u>82</u> – ( 82 – $A_{cl}$) nucleons, where $A_{cl}$ is the mass number of the cluster (fig.1). In the third step, the fragment pair created in the second step evolves towards scission in spite of a strong confinement due to the own Coulomb barrier of the two fragments [8].

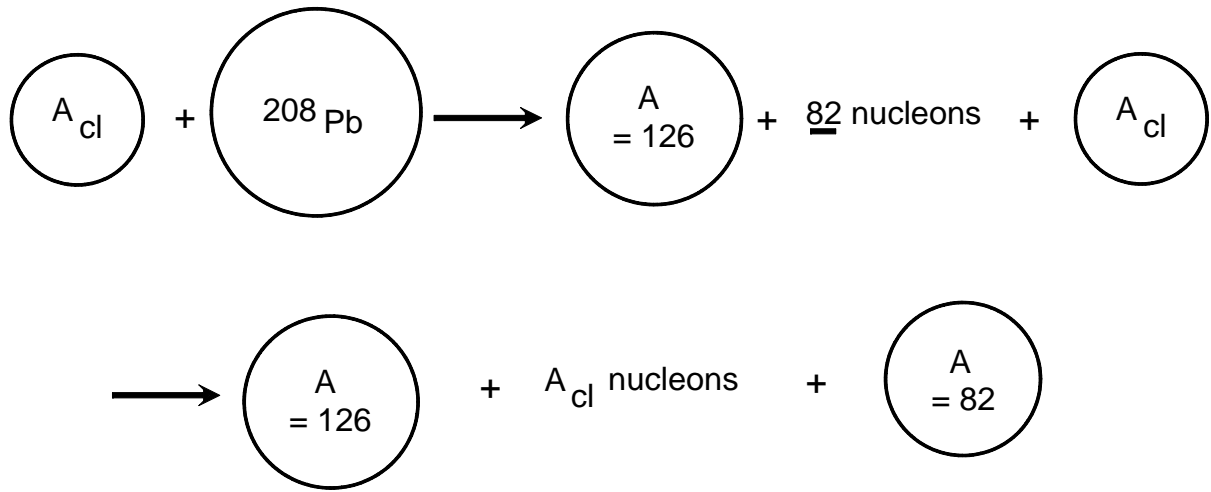

**Fig.1** *Formation of the nascent light and heavy fission fragments made of an A = 82- and an A = 126-nucleon core.*

In the nucleon-phase model, during the second step of 0.17 yoctosecond [6], a new state of nuclear matter is formed, in which nucleon shells, instead of proton- and neutron- shells, are closed at "magic mass numbers", 82 and 126 [9].

The nucleon-phase model explains not only the asymmetric mass distributions of fission fragments, but also the symmetric distributions [10].

The law of Flynn et al. [11] proves that an *on average* constant number of nucleons is transferred from the $^{208}$Pb core to the primordial cluster in the formation of the light fragment [12]. The modern expression of the law is that the *mean mass* of the light fragment is given by

$$\overline{A_L} = A_{cl} + 68 \ [13] \tag{2}$$

Fig.2 shows how happens, on average, the formation of the light fragment.

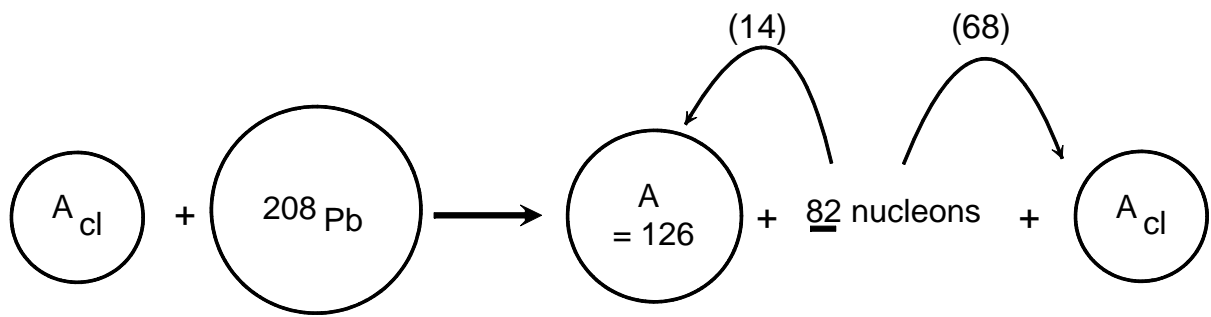

**Fig.2** : *Distribution of the 82 nucleons released in the destruction of the lead 208-core between the A =126 nucleon-core and the primordial cluster of mass number $A_{cl}$, according to the law of Flynn et al. revisited.*

## 3. Justification of the existence of a new mode of ternary fission.

It may be asked: "What really happens when 82 nucleons are released in the destruction of the $^{208}$Pb core?" Most of them are certainly captured by the primordial cluster; and most of the remaining nucleons are certainly shared between the nascent light and heavy fragments, made of an A = 82 nucleon core and of an A = 126 nucleon core, respectively, as shown schematically in fig.1.

But what happens to the 82 free nucleons if they are accidentally separated from the primordial cluster by the A = 126 nucleon core, as shown in fig.3-a, where the 82 nucleons cannot "see" the cluster because it is screened by the A = 126 core?

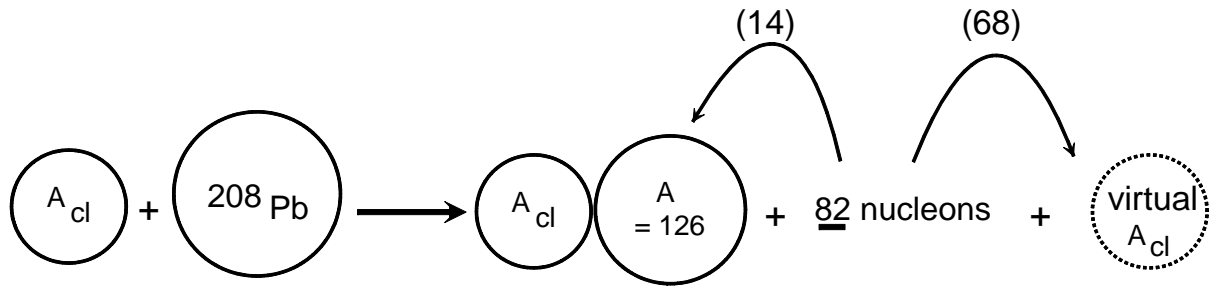

**Fig.3-a** : *A relative position of primordial cluster, A = 126 nucleon core and free nucleons in which 68 nucleons, on average, could escape capture by the cluster.*

Let us *assume* that this law (eq.2) still holds in the accidental situation, and that consequently , at the end of the nucleon phase, the 68 nucleons, which are not captured by the cluster, nevertheless aggregate into a new "cluster", on average an A = 68 cluster.

In this situation, the final configuration could , on average, be the following:

$A_{cl}$ _______ $\overline{A_H}$ =140 _______ $\overline{A_L}$ = 68

i.e. a ternary *colinear* configuration (fig.3-b), which must be "*colinear*" because it results from the accidental situation that $A_{cl}$ was "screened" by the A = 126 core .

According to this hypothesis, any particular fission event occurring in such an accidental situation should, *in fact*, lead to the creation of a light fragment of mass going, in principle, from zero to 82, but more probably *near to A = 68,* and of a heavy fragment of mass going, in principle, from 126 to 208, but more probably *near A = 140*, but on condition that $A_L$ + $A_H$ is still equal to 208.

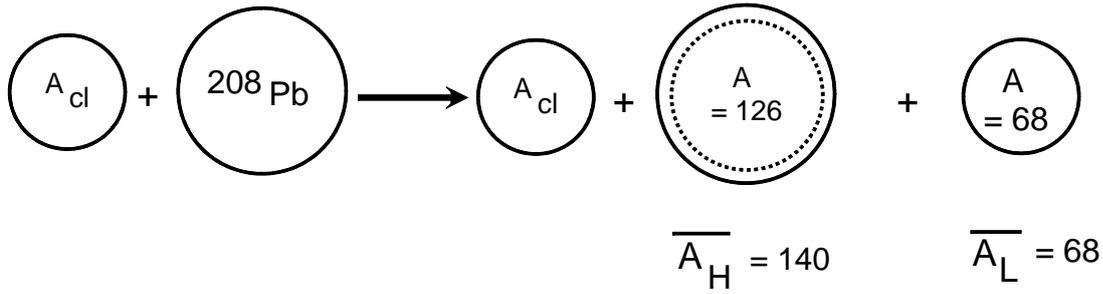

$$\overline{A_H} = 140 \qquad \overline{A_L} = 68$$

**Fig.3-b** : *In the accidental relative position of fig.3-a, a light fragment made , on average, of 68 nucleons could be formed by aggregation.*

Let us now consider in particular the spontaneously fissioning $^{252}$Cf nucleus.

Its "clusterized" primordial configuration is $^{208}$Pb—$^{44}$S, and its clusterization energy is equal to 106.90± 0.40 MeV [14]. But the $^{208}$Pb core, with its binding energy per nucleon of only 7.867.45 keV, is easily destroyed in a collision with the $^{44}$S-cluster ($E_B/A$ = 7 994 ± 9 keV) [14].

Among the possible fragment pairs resulting from the redistribution of the <u>82</u> free nucleons, let us consider the most energy-rich Sn/Ge fragment pair, the $^{128}$Sn-$^{80}$Ge pair (see tab.1), with its internal energy Q = 131.10 MeV±0.06 MeV [14]. This pair forms with the $^{44}$S primordial cluster of $^{252}$Cf (s.f.) the colinear system:

$$^{44}\text{S} \;\underline{\qquad}\; ^{128}\text{Sn} \;\underline{\qquad}\; ^{80}\text{Ge} \;. \qquad\qquad (3)$$

| $Z_H/Z_L$ | Fragment pair | Released energy Q (MeV) [14] |
|---|---|---|
| | | |
| In/As | $^{125}$In – $^{83}$As | 128.61 (0.25) |
| | $^{127}$In -$^{81}$As | 127.76 (0.05) |
| | $^{126}$In- $^{82}$As | 126.38 (0.24) |
| | | |
| Sn/ Ge | $^{128}$Sn – $^{80}$Ge | 131.099 (0.060) |
| | $^{130}$Sn-$^{78}$Ge | 130.25 (0.02) |
| | $^{126}$Sn- $^{82}$Ge | 129.89 (0.26) |
| | $^{129}$Sn-$^{79}$Ge | 128.33 (0.12) |
| | | |
| Sb/Ga | $^{131}$Sb-$^{77}$Ga | 126.231 (0.025) |
| | $^{133}$Sb-$^{75}$Ga | 125.66 (0.03) |
| | | |
| Xe/Ni | $^{136}$Xe-$^{72}$Ni | 118.60 (0.44) |
| | $^{138}$Xe-$^{70}$Ni | 117.53 (0.39) |

**Table 1** : Energy released by the dissociation of $^{208}$Pb into various energy-rich fragment pairs

The internal energy of this system, made of the sum of Q and of the clusterization energy of $^{252}$Cf, 106.90 MeV, is equal to 238.0 $\pm$ 0.5 MeV. It is a considerable energy.

The Coulomb barrier $B_c$ of the Sn/Ge pair is equal to only 166.52 MeV; and that of the Sn/S pair is equal to only 90.82 MeV. Even if a sphericity correction of about 11 MeV, at the most [8], was added to $B_c$ (because the Sn nucleus is more spherical), the internal energy of this energy-rich, axially vibrating system should be great enough in comparison with the Coulomb barrier for escaping its confinement.

## 4. Comparison with the results of Pyatkov et al.

Thanks to two ingenious experimental devices, Yu. V. Pyatkov et al. succeeded in showing that *two emitted particles originate from an A = 208 body*, both in $^{252}$Cf (s.f.) and in $^{235}$U + n$_{th}$, as shown in fig. 7-c of ref.[1], and further in showing that *the most probable* $M_1$, $M_2$ *masses are 68, 139* , as shown in their fig. 6a for the $^{252}$Cf nucleus and for a sum $M_1+M_2$ equal to 208 [1]. Moreover, these authors report the observation in $^{252}$Cf of a "cluster" $^{44}$S, i.e. of the "primordial cluster" of $^{252}$Cf.

These observations are in perfect agreement with the predictions of Sect.3. Even the observation of light clusters as heavy as $^{48-56}$Ca is not in contradiction with these predictions, as will be shown in Sect 5.

## 5. Discussion

In Sect.3, we have assumed that the new expression of the law of Flynn et al., i.e.

$$\overline{A_L} = A_{cl} + 68,$$

might still hold *when the 82 free nucleons* can still have access to the A= 126 nucleon core but can't have access to the primordial cluster any longer.

This assumption might appear very surprising at first sight.

However, the low yield of the ternary emission process, 4-5 10$^{-3}$ of the binary fission yield according to Pyatkov et al. clearly indicates that this emission is exceptional. It is the first point.

Moreover, even if the lifetime of the nucleon phase, during which the light fragments are formed, appears infinitesimally small, it might nevertheless be great enough for permitting small changes in the geometry of the ternary system made of $^{44}$S, 82 free nucleons and A = 126 nucleon core: For example, the primordial cluster $^{44}$S could *escape* from its usually freely accessible position and become shielded from the 82 free nucleons by the A = 126 nucleon core (fig.3- a). It is a second point.

Anyway, the nucleons which are not captured by the A = 126 nucleon core can still *aggregate into a new "cluster"*, as do the valence nucleons of the $^{208}$Pb core of any fissile nucleus [7] (fig.3-b). It is a third point.

Concerning the observation of light clusters as heavy as $^{48-56}$Ca, the fact that up to 12 nucleons are captured by the primordial cluster $^{44}$S of $^{252}$Cf does not contradict that the A= 126 nucleon core plays the role of  a *screen*, since without this screen the $^{44}$S cluster would have captured on average 68 of the 82 nucleons released by the destruction of $^{208}$Pb.

## 6. Conclusion

The discovery of the  colinear ternary fission by Yu.V. Pyatkov et al. constitutes a great advance in our understanding of the phenomenon of nuclear fission.

It confirms the role played by the primordial cluster, which here plays a role of spectator in the scenario of the dissociation of the primordial core. It also confirms the eminent role of the dissociation of this $^{208}$Pb core into cores having a magic *mass*- number, in particular into the A = 126 nucleon core.

For all these reasons, the discovery of  Yu. V. Pyatkov et al. can be described in the framework of the nucleon phase model of nuclear fission, and, in our opinion, it constitutes a necessary extension of this model.

This new ternary fission mode is distinct from the other ternary fission modes, the "orthogonal" emission mode and "the isotropic" emission mode [15].

June 29,2010.